\documentclass[a4paper,11pt]{article}
\pdfoutput=1 

\usepackage{jcappub} 

\usepackage[T1]{fontenc} 

\usepackage{afterpage}

\def\Neff{N_{\rm eff}}
\def\mster{m_{\nu, \rm sterile}^{\rm eff}}
\def\dm2{\Delta m_{41}^2}
\def\sinsq{\sin^22\theta}
\def\sinsqe{\sin^22\theta_{14}}
\def\sinsqmt{\sin^22\theta_{24}}
\def\LCDM{$\Lambda$CDM}
\def\ket[#1]{|#1\rangle}
\newcommand\Tstrut{\rule{0pt}{2.6ex}}         
\newcommand\Bstrut{\rule[-0.9ex]{0pt}{0pt}}   

\usepackage{array}
\usepackage{subfigure}
\newcolumntype{C}[1]{>{\centering\arraybackslash}p{#1}}
\def\leaderfi1{\leaders\hbox to 5pt{\hss.\hss}\hfil}
\newenvironment{rcases}
  {\left.\begin{aligned}}
  {\end{aligned}\right\rbrace}

\title{\boldmath Cosmological constraints on sterile neutrino oscillations from \emph{Planck}}

\author[1]{Alan M. Knee,}
\author[2,3]{Dagoberto Contreras,}
\author[1]{Douglas Scott}

\affiliation[1]{Department of Physics \& Astronomy\\
University of British Columbia, Vancouver, BC, V6T 1Z1  Canada}
\affiliation[2]{Department of Physics \& Astronomy\\
University of York, Toronto, ON, M3J 1P3  Canada}
\affiliation[3]{Perimeter Institute for Theoretical Physics, Waterloo, ON, N2L 2Y5 Canada}

\emailAdd{alan.knee@alumni.ubc.ca}
\emailAdd{dago@yorku.ca}
\emailAdd{dscott@phas.ubc.ca}

\abstract{Both particle physics experiments and cosmological surveys can
constrain the properties of sterile neutrinos, but do so with different
parameterizations that naturally use different prior information. We present
joint constraints on the 3$+$1 sterile neutrino model oscillation parameters,
$\dm2$ and $\sinsq$, with log priors on those parameters using mostly
cosmological data from the \emph{Planck} satellite. Two cases are
considered, one where the sterile neutrino mixes with electron neutrinos
solely, and another where the sterile neutrino mixes exclusively with muon
neutrinos, allowing us to constrain the mixing angles $\sinsqe$ and $\sinsqmt$, along with the squared mass-splitting $\dm2$.
We find that cosmological data are inconsistent with strong hints of a sterile neutrino coming from some oscillation channels of the LSND and MiniBooNE experiments, under the assumption that the sterile neutrinos mix with a single neutrino flavour.
We also forecast the sensitivity with which future CMB experiments should be able to
probe $\dm2$ and $\sinsq$ in log space.}

\notoc

\begin{document}
\maketitle
\flushbottom

\newcommand{\pz}{\phantom{0}}

\date{\today}

\pagebreak


\section{Introduction}
\label{sec:intro}

The neutrino sector is still not well understood. In particular the masses and
mixing angles of the three Standard Model (SM) neutrinos are still being determined,
along with the possibility that there might be additional light leptons.
Cosmology has an important role to play here, since the constraints obtained
are quite complementary to those coming from particle experiments (see the
Review of Particle Physics articles on neutrinos in cosmology and neutrino
oscillations in ref.~\cite{t}). Models involving more than three types of
neutrino have mostly been disfavoured by data from both particle physics and
cosmology experiments. However, some neutrino experiments have hinted
at the existence of additional neutrino species, and it is often hypothesized that these new species, should they exist, are ``sterile'' neutrinos. Such
neutrinos differ from the three standard ``active'' neutrinos in that they are
not charged under the weak interaction and thus lack most of the fundamental
interactions experienced by the active neutrinos in the SM.
Instead, sterile neutrinos only interact in the cosmological context via gravitation and through
mixing with the other neutrinos \cite{q}.

The study of sterile neutrinos presents an interesting avenue into new physics
beyond the SM. There are many different experiments whose results have hinted
at evidence for non-trivial content in the neutrino sector. In perhaps the most
prominent example, results from LSND are in tension with other
experiments, since the standard 3-neutrino model does not adequately match
the data \cite{i}. The results are well fit with an eV sterile neutrino in a
3+1 neutrino model, although other experiments measuring the same channel exclude
this model (see for example ref.~\cite{Agafonova:2018dkb}). Surprisingly, this
anomalous result appears to have been recently corroborated by MiniBooNE,\footnote{Though not
consistently for all data channels, see ref.~\cite{b}.} which, combined with the
data from LSND, yields an electron excess at the level of $>6\sigma$
significance \cite{b}. At the present moment the union of all oscillation
experimental results seems to be simultaneously at odds with the standard 3-neutrino model 
and a 3+1 sterile neutrino model, even although some data prefer one
over the other. This provides motivation for independently considering the 3+1
sterile neutrino model from a cosmological perspective.

This paper is structured as follows: In section~\ref{sec:nucosmology} we review
the effect that neutrinos have on cosmology and specifically the CMB.
Section~\ref{sec:model} introduces the specific sterile neutrino picture that we
constrain here. We present our constraints on existing cosmological data in
section~\ref{sec:results}. We forecast how well constraints for a future survey
with similar capabilities to the CMB ``Stage 4'' experiment will perform in section~\ref{sec:forecast}.
Finally we conclude in section~\ref{sec:conclusions}.

\section{Neutrinos in cosmology and particle physics}
\label{sec:nucosmology}

Neutrino physics is constrained by the cosmic microwave background (CMB) in
terms of two parameters, $\Neff$ and $\Sigma m_\nu$. The quantity $\Neff$ is
the effective number of neutrino species, and \emph{Planck} data are consistent
with $\Neff=3.046$, as predicted by the SM. The SM prediction differs slightly
from $\Neff=3$ due to heating of the neutrinos by electron-positron
annihilation during the epoch of neutrino decoupling (see refs.~\cite{j, r, x}
for discussion of how the standard value of 3.046 is obtained). Large
deviations from this value, on the order of unity, are ruled out by
\emph{Planck} at over 99\% confidence, but the constraints do allow for small
deviations from 3.046 \cite{c}. The second cosmologically relevant quantity,
$\Sigma m_\nu$, is the sum of the active neutrino masses. Particle physics
experiments have only been able to measure the mass-squared differences between
neutrino mass states,\footnote{In terms of the mass of neutrinos, measurements of an
effective electron neutrino mass are also possible; nevertheless, the current cosmological
constraints have little distinguishing power with respect to the neutrino
hierarchy.} so we have the freedom to consider different mass hierarchies.
Usually, for cosmology, one assumes the normal hierarchy, which consists of two
very light neutrinos, $m_1$ and $m_2$, with the former being
the lightest, and a heavier neutrino, $m_3$; however, inverted or degenerate
hierarchies are also possible. We will assume the normal hierarchy here
(having an inverted hierarchy would strengthen \emph{Planck} constraints on
sterile neutrinos), in which there are two nearly massless neutrinos and one massive neutrino
with $\Sigma m_\nu=0.06$ eV \cite{c}. An approximation that is reasonable for
cosmological purposes is to take $m_1$ and $m_2$ to be massless, leaving
$\Sigma m_\nu$ to be dominated by $m_3$. When one adopts a model that allows
for sterile neutrinos, $\Neff$ is reinterpreted so that deviations from 3.046
(i.e. $\Delta\Neff=\Neff-3.046$) indicate an additional degree of freedom
associated with the sterile neutrinos \cite{s}.

SM neutrinos contribute a cosmological density parameter given by \cite{u}
\begin{equation}
  \label{eq:omeganu}
  \Omega_\nu=\frac{\rho_\nu}{\rho_{\rm crit}}\approx \frac{\sum c_i g_i^{3/4}
  m_i}{94.1\,h^2\,{\rm eV}}\approx \frac{\sum m_\nu}{93.03\,h^2\,{\rm eV}},
\end{equation}
where $h\equiv H_0/(100\,\rm{km}\,\rm{s}^{-1}\,\rm{Mpc}^{-1})$ is the reduced
Hubble constant. In cosmology, 
only the three SM neutrinos are included in the mass sum $\sum m_{\nu}$.
The masses $m_i$ represent the active mass eigenstates, $c_i$
is the degeneracy per state, and $g_i=3.046/3$ is a degeneracy factor that
arises because we have $\Neff=3.046$ distributed among three neutrinos. The
three-quarters power comes from the translation between a number density and an
energy density within the calculations, and will be elaborated on in section \ref{sec:conversion}. We stress that this relation is only valid in the instantaneous
decoupling limit, meaning that the neutrinos are perfectly Fermi-Dirac
distributed. In reality, this decoupling is not instantaneous, and heating from
electron-positron annihilation, in addition to other quantum-mechanical effects
in the early Universe, causes the neutrinos to no longer be exactly Fermi-Dirac,
yielding a slightly different factor in the denominator \cite{t, r}:
$\Omega_{\nu}= \Sigma m_\nu/93.14\,h^2\,{\rm eV}$. To be consistent with the
\emph{Planck} analysis, we have opted to use the parameterization in
eq.~\ref{eq:omeganu}. It is worth pointing out that this means that our
approach is not entirely self-consistent, since it is this departure from being
perfectly Fermi-Dirac that leads to $\Neff=3.046$ instead of $\Neff=3$ in the
first place. However, this difference in the second decimal digit is small
enough that it will not significantly influence our results.

Sterile neutrinos introduce an additional energy density to the cosmological
background, which we can parameterize with $\Omega_{\nu, \rm sterile}$ as in
the \emph{Planck} parameters paper \cite{c}:
\begin{equation}
\mster\equiv 94.1\,\Omega_{\nu,\rm sterile}h^2\,\rm{eV}.
\end{equation}
In this case, the total physical neutrino density becomes $\Omega_\nu h^2 =
0.00064 + \Omega_{\nu, {\rm sterile}}h^2$, where 0.00064 is the contribution to
the physical neutrino density by the three active neutrinos, assuming the
standard mass sum $\Sigma m_\nu = 0.06$ eV. Thus, any change from $\Omega_\nu
h^2=0.00064$ is assumed to be due to sterile neutrinos. This picture is for the
simplest case where the sterile neutrinos do not couple with any other
neutrino, hence the factor of 94.1 eV.

In particle physics, neutrino properties are mostly constrained through
oscillation experiments. The exact masses of neutrinos are unknown, and
neutrino experiments instead probe the mass-squared differences $\Delta
m_{ij}^2$ between two mass eigenstates. For the study of sterile neutrinos we
parameterize our model in terms of $\dm2$, the mass-squared difference between
$m_4$ and the lightest mass state. In the normal hierarchy with $m_1=0$, this
reduces to a convenient equality: $\dm2=m_4^2-m_1^2=m_4^2$. Sterile neutrinos,
if they exist, are usually considered to be significantly more massive than the
three standard species, with a mass closer to the eV scale \cite{m}, which has
implications for cosmology \cite{q}.

For neutrinos, mass states do not correspond uniquely to the flavour states.
Instead, the two bases are related through a mixing
matrix $U_{\alpha \beta}$. Interpreted in the 3+1 scenario, the mixing matrix
is rectangular with $\alpha=e,\mu,\tau$, and $\beta=1,2,3,4$. The elements of
the mixing matrix are products of sines and cosines of the mixing angles
$\theta_{ij}$ ($i,j=1,2,3,4$). The flavour states are thus linear combinations
of the mass states and vice versa:
\begin{equation}
\ket[\nu_\alpha]=\sum_\beta U_{\alpha \beta}^*\ket[\nu_\beta].
\end{equation}
Finding stringent constraints on these mixing angles remains an active field of
research in experimental particle physics \cite{Neutrino2012}. Neutrino oscillation experiments
measure the flux of neutrinos prepared in a preferred flavour at two detectors
separated by some large distance, $L$.
The $\nu_e$ and $\nu_\mu$ disappearance searches are concerned with measuring
how many neutrinos change flavour as a result of neutrino oscillation, whereas
$\nu_\mu \rightarrow \nu_e$ appearance searches measure how many electron neutrinos
appear from an initial beam of muon neutrinos. Sterile neutrinos could mix
with one or more neutrino flavours; however, in this paper we will only be
considering the simpler cases where they mix with a single active species.
The survival probabilities in $\nu_e$ and
$\nu_\mu$ disappearance searches are, respectively, given in natural units by \cite{m}
\begin{align}
P^{3+1}_{ee}&=1-4|U_{e4}|^2(1-|U_{e4}|^2)\sin^2\bigg(\frac{\dm2 L}{4E}\bigg)=1-\sin^22\theta_{ee}\sin^2\bigg(\frac{\dm2 L}{4E}\bigg), \\
P^{3+1}_{\mu\mu}&=1-4|U_{\mu4}|^2(1-|U_{\mu4}|^2)\sin^2\bigg(\frac{\dm2 L}{4E}\bigg)=1-\sin^22\theta_{\mu\mu}\sin^2\bigg(\frac{\dm2 L}{4E}\bigg),
\end{align}
where $L$ is the beam length, $E$ is the energy of the neutrinos, and
\begin{align}
\sin^22\theta_{ee}&=4|U_{e4}|^2(1-|U_{e4}|^2), \\
\sin^22\theta_{\mu\mu}&=4|U_{\mu4}|^2(1-|U_{\mu4}|^2).
\end{align}
Using the specific parameterization of the mixing matrix in ref.~\cite{m},
we have
\begin{align}
|U_{e4}|&=\sin\theta_{14}, \\
|U_{\mu4}|&=\cos\theta_{14}\sin\theta_{24}.
\end{align}
One therefore finds that, in the case where the sterile neutrino only mixes
with a muon neutrino (so $\cos\theta_{14}=1$), the effective sterile neutrino
mixing angle reduces to $\sin^22\theta_{24}=\sin^22\theta_{\mu\mu}$, and
if it only mixes with an electron neutrino the mixing is instead
determined by $\sin^22\theta_{14}=\sin^22\theta_{ee}$. Thus, in both
situations there is
a direct correspondence between sterile neutrino oscillation and either
$\nu_e$ or $\nu_\mu$ disappearance, depending on which flavour the
sterile neutrino mixes with. We assume $\sin^22\theta_{34}=0$ at all times, in accordance with experimental measurements. More
complicated models could in principle have a sterile neutrino that mixes
with multiple species, or have more than one sterile species. See
ref.~\cite{m} for a review covering multiple mixing cases and the
2-sterile neutrino (3+2) scenario. In particular, for $\nu_e$ appearance in $\nu_\mu$ beams the probability amplitude is given by \cite{m}
\begin{equation}
\sin^22\theta_{\mu e}=4|U_{\mu 4}U_{e4}|^2=\sin^2\theta_{24}\sin^22\theta_{14}.
\end{equation}
This requires a model in which the sterile neutrino mixes with more than one neutrino flavour, since this mixing angle is zero if either $\theta_{14}$ or $\theta_{24}$ are zero. Hence, the results we present here will only be directly comparable with particle physics constraints from disappearance channels, as they will have used the same assumptions as we have.


\section{Model}
\label{sec:model}

The main complication in comparing cosmological constraints with oscillation
measurements arises from the fact that cosmology and neutrino experiments are
sensitive to different physical effects and so focus on different
parameterizations and priors. In particular, flat priors in one parameter space
will not be flat in a different space. We will describe this in general in
section~\ref{sec:priors}. In refs.~\cite{d,k}, a method is developed to express
the $(\Neff,\mster)$ parameter space in terms of the $(\sinsq,\dm2)$ parameter
space and vice versa, which was used in ref.~\cite{a}, allowing constraints to be
directly compared on sterile neutrinos from both the CMB and neutrino-oscillation 
experiments by converting between parameter spaces. We will explain
this conversion in detail in section~\ref{sec:conversion}. Throughout this
paper, we will refer to these two spaces respectively as the ``cosmology'' and
``particle'' parameter spaces. Our particular goal is to elaborate on the results of ref.~\cite{a}
by using a model that has uninformative priors in the particle parameter space.
We constrain the particle parameters by way of a Markov chain Monte Carlo
analysis of \emph{Planck} data using \texttt{CosmoMC} \cite{e,f}, as well as
forecasting how well future experiments could improve upon these constraints.
Our cosmological model includes the base-\LCDM\ parameters, as well as $\Neff$,
and $\mster$, but with a non-flat prior on the cosmology parameters, such that
the particle space parameters are initialized with flat priors, which we now describe in more detail.

\subsection{Priors}
\label{sec:priors}

Recall that the purpose of a Markov chain Monte Carlo is to produce a random
sample from a distribution $p(\alpha|X)$ known as the posterior, where $\alpha$
is a model parameter and $X$ represents the data. It follows from Bayes'
theorem that
\begin{equation}
  \label{eq:prior}
  p(\alpha|X)=\dfrac{p(X|\alpha)p(\alpha)}{p(X)} \propto \mathcal{L}(X|\alpha)p(\alpha),
\end{equation}
where $\mathcal{L}(X|\alpha)=p(X|\alpha)$ is the likelihood function and
$p(\alpha)$ is the prior, representing an initial belief about the parameter
distributions. The posterior is thus proportional to the product of the
likelihood and the prior, a relation that forms the basis of Bayesian
inference. As the name suggests, the prior represents the state of our knowledge
of a parameter \emph{before} considering specific data. Picking a prior is subjective,
but it is often chosen to be uninformative, that is it gives no information on the
value of the parameter and thus represents our ignorance~\cite{Gregorybook}. In
our case we will also be influenced by trying to be consistent with oscillation
experiments.
In our MCMC calculations, we want to include the particle parameters as model
parameters, with flat priors in logarithmic space, varying them by proxy as we vary $\Neff$ and
$\mster$ within \texttt{CAMB} \cite{y}. However, a uniform distribution in one
set of parameters is generally not uniform when transforming to a different
set, and thus simply varying $\Neff$ and $\mster$ with flat priors and then
converting them to the particle space would not yield flat priors in the
particle space.

The prior in one parameter space that gives a uniform prior in another space is
given by the Jacobian relating the two parameter spaces. This directly follows
from the fact that given two probability distributions, e.g.~$p_{\rm
cosm}(\Neff, \mster)$ and $p_{\rm part}(\log(\sinsq), \log(\dm2))$, related to each other
by a change of variables transformation, the following relation holds:
\begin{equation}
  \label{eq:4}
  p_{\rm cosm}(\Neff, \mster)\,d(\Neff)\,d(\mster)=p_{\rm part}(\log(\sinsq,
  \dm2))\,d(\log(\sinsq))\,d(\log(\dm2)).
\end{equation}
Therefore if one desires a logarithmic prior, $p_{\rm part}(\log(\sinsq), \log(\dm2))={\rm
constant}$, on the particle parameters, the prior for the cosmology parameters
must satisfy
\begin{equation}
p_{\rm cosm}(\Neff, \mster) \propto \biggl|\biggl|\frac{\partial(\log(\sinsq))\,\partial(\log(\dm2))}{\partial(\Neff)\,\partial(\mster)}\biggl|\biggl|,
\end{equation}
which is the Jacobian determinant for this change of variables.
\begin{figure}[h]
\centerline{\includegraphics[width=1\textwidth]{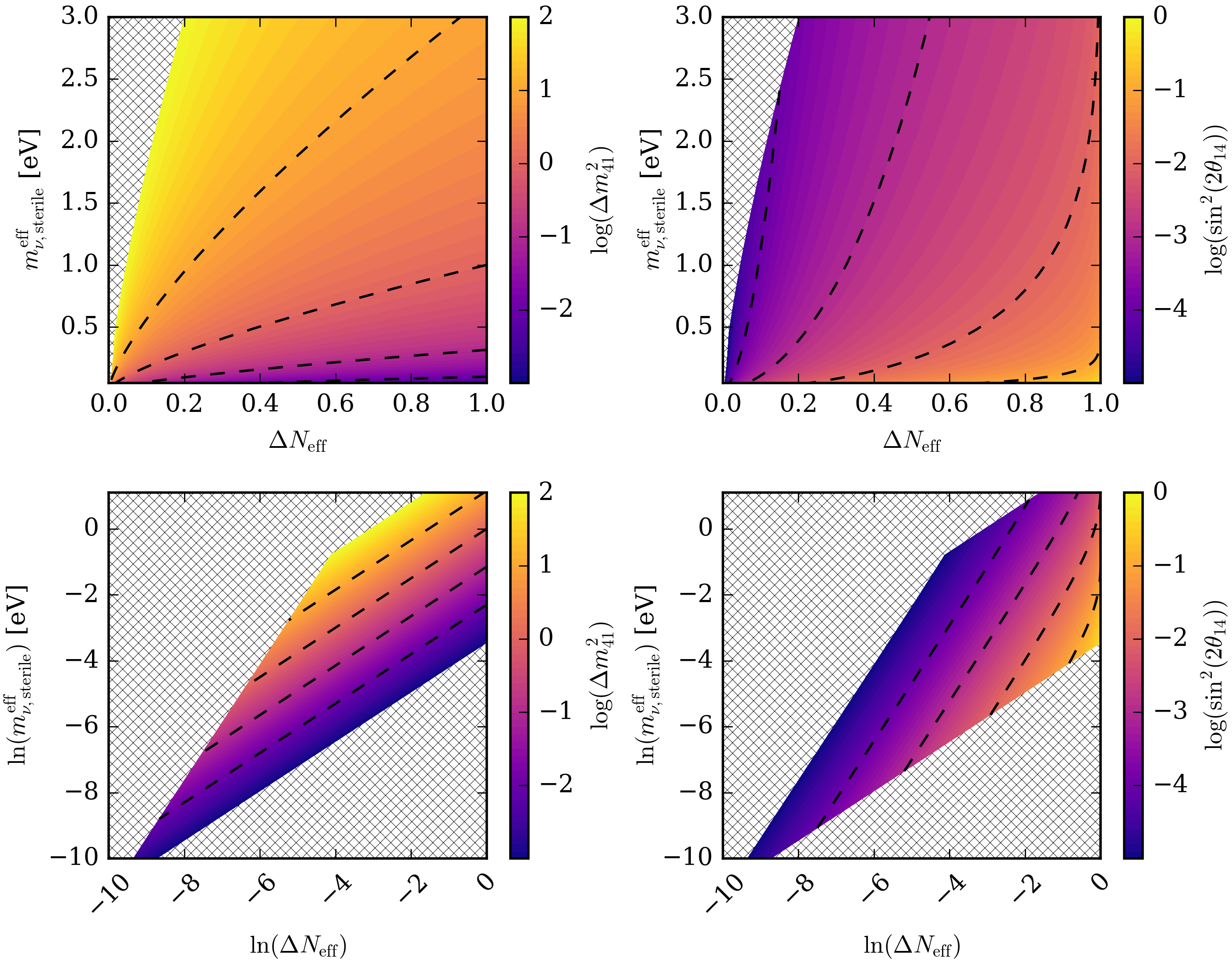}}
\caption{Plots of the \texttt{LASAGNA} output in linear and log spaces of $\Delta\Neff$ and $\mster$, in which the sterile neutrino mixes with an electron neutrino. The blue regions in the pair of logarithmic plots (lower panels) only intersect with each other at extremely small values of $\Delta\Neff$ and $\mster$. In linear space, the region of overlap is not even visible and thus sampling it properly is challenging. In log space, this region is greatly expanded and can be well sampled. The hatched regions are excluded by the choice of priors (eq.~\ref{eq:parrange}). This same set of plots for the case where the sterile neutrino mixes with a muon neutrino looks very similar.}
\label{fig:paramspace}
\end{figure}

\subsection{Parameter-space conversions} \label{paramspaceconv}
\label{sec:conversion}

Since the particle parameters are not included in the base
\texttt{CosmoMC} program,
we instead vary $\Neff$ and $\mster$ and compute the particle parameters
from the chains. To accomplish this, we follow the procedure described
in ref.~\cite{a} to compute the cosmology parameters on a grid of values for
$\dm2$ and $\sinsq$ in the ranges
\begin{equation}
\label{eq:parrange}
10^{-3}\leq\dm2\leq10^{2}\,\,\rm{eV}^2, \quad 10^{-5}\leq\sinsq\leq1.
\end{equation}
The upper prior limit of 100 eV$^2$ for the mass splitting is consistent
with setting an upper prior limit of $m_4<10$ eV on the physical sterile
neutrino mass, as done in the Planck Collaboration analysis \cite{c}, which is chosen
because at masses greater than 10 eV the sterile neutrinos begin to behave
like cold dark matter and there is no need to handle them separately.
For each point on the grid, we use the code \texttt{LASAGNA} \cite{d}
to compute $\Neff$ and $\mster$, yielding a mapping between the two
parameter spaces.
\texttt{LASAGNA} solves the quantum kinetic equations for sterile neutrino
thermalization in the temperature range $1\leq T\leq 40$ MeV, generating
discrete values of $x=p/T$ and $P_{\rm sterile}^+$, where $p$ is the
neutrino momentum, $T$ is the temperature, and $P_{\rm sterile}^+$ is
defined as
\begin{equation}
P_{\rm sterile}^+=(P_0+\bar{P_0})+(P_z+\bar{P_z}),
\end{equation}
where $P_i$ are the elements of the quantum state vector ${\bf P}=(P_0, P_x, P_y, P_z)$ for a particular momentum mode.
This output is used to evaluate the integral expression
\begin{equation}
\Delta\Neff=\Neff-3.046=\frac{\int dx\, x^3 (1+e^x)^{-1} P_{\rm
sterile}^+}{4\int dx\, x^3 (1+e^x)^{-1}}
\end{equation}
by summation. Refs.~\cite{d, k} provide a detailed
explanation of the formalism behind the \texttt{LASAGNA} code. To obtain
an expression for $\mster$, we assume thermally-distributed
sterile neutrinos about some temperature $T_{\rm sterile}$ \cite{a, c}:
\begin{equation}
\Delta\Neff=(T_{\rm sterile}/{T_\nu})^4.
\end{equation}
When $T_\nu=T_{\rm sterile}$, the sterile neutrinos thermalize at the same
temperature as the active neutrinos, resulting in full thermalization of
the sterile neutrinos with $\Delta\Neff=1$. The $T^4$ proportionality comes
from the fact that $\Neff$ is parameterizing an energy density.
Matter densities, on the other hand, are proportional to $T^3$. Using $\dm2=m_4^2$, we can compute $\mster$ directly:
\begin{align}
\mster&=(T_{\rm sterile}/{T_\nu})^3m_4^{\rm thermal} \\
&=(\Delta\Neff)^{3/4}m_4^{\rm thermal} \\
&=(\Delta\Neff)^{3/4}\sqrt{\dm2}.
\label{eq:1}
\end{align}
\begin{figure}[h]
\centering
\includegraphics[width=0.45\textwidth]{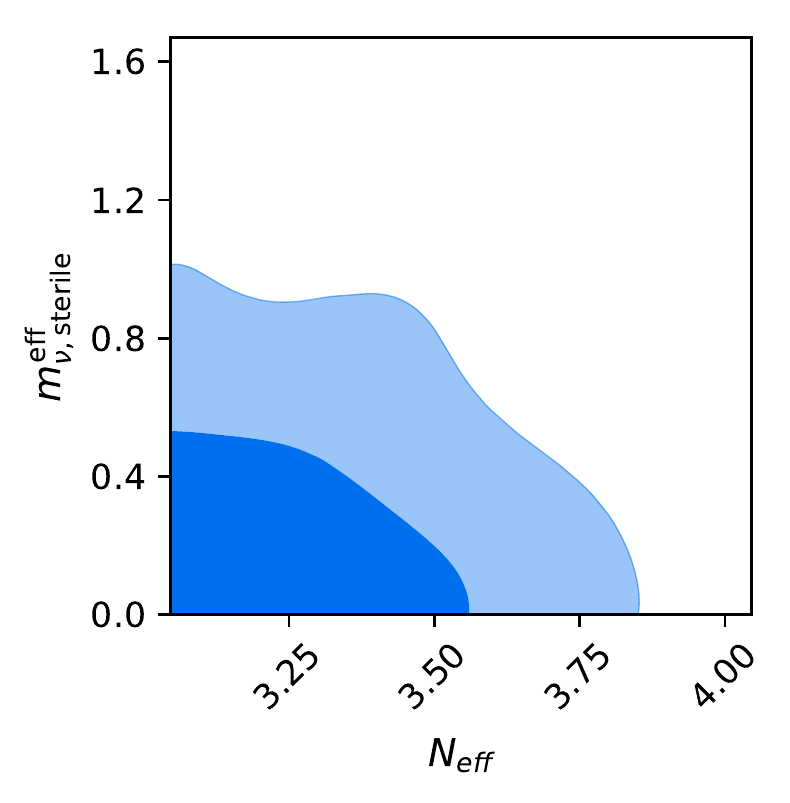}
\includegraphics[width=0.45\textwidth]{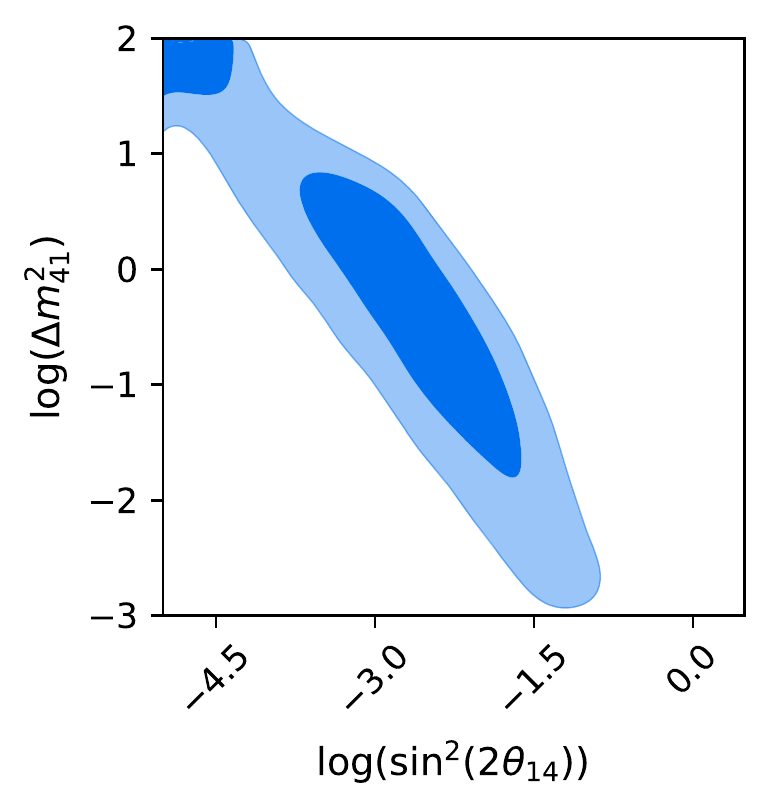}
\caption{\emph{Left panel}: \emph{Planck} 68\% and 95\% CL constraints in the cosmology parameter space using \emph{Planck} CMB TT+lowTEB data. Priors are flat in the ranges $3.046 \leq \Neff \leq 4.046$ and $0 \leq \mster \leq 3$ eV. \emph{Right panel}: The procedure explained in section \ref{paramspaceconv} is used to calculate the particle parameters from the original cosmology chains. The region of low mass and mixing angle appears to be ruled out, but these constraints are not robust, since this is a result of poor sampling. A solution to this is discussed in section \ref{sampling}. Note that here $\log x$ denotes the base 10 logarithm, not the natural logarithm, which we will always write as $\ln x$.}
\label{fig:planck_direct}
\end{figure}

The \texttt{LASAGNA} calculations are quite sensitive to the initial lepton
asymmetry $L$, which we take to be zero. The effects of varying this
asymmetry have been studied previously \cite{a}, though for simplicity we do not
consider it here.
Such models may offer additional avenues for studying sterile neutrinos in
cosmology. For example, it has been suggested that a large asymmetry could
suppress the thermalization of sterile neutrinos in the early Universe, thereby
allowing sterile neutrinos to have a large mixing angle without requiring
$\Delta\Neff\approx1$, i.e. complete thermalization \cite{k, w, v}.

\texttt{LASAGNA} supports a single sterile neutrino species that mixes with one
active neutrino species. The flavour of the active species can be either an
electron neutrino or a muon neutrino, corresponding to the mixing angles
$\theta_{14}$ and $\theta_{24}$, respectively \cite{m}. We thus have two
different cases, each parameterized by a different mixing angle. Both cases are
considered in our MCMC runs, allowing us to obtain constraints on either
$\sinsqe$ or $\sinsqmt$.

\subsection{Sampling} \label{sampling}

In order to avoid a false detection of sterile neutrinos, it is of
critical importance to ensure that the region of parameter space
corresponding to small mixing angles and mass splittings is well sampled.
Indeed, if we simply compute the particle
parameters from $\Neff$ and $\mster$, as was done in figure
\ref{fig:planck_direct}, the region of low mass and mixing angle appears to
be ruled out at 95\% confidence. However, this is not a genuine detection of high mass,
but rather a consequence of the chains being unable to sample from this
region. In figure \ref{fig:paramspace}, we can see that the region of
particle space where both the mixing
angle and mass splitting are small is where the darker regions overlap between the two panels. In
the cosmology parameter space, there is practically no overlap at all and
thus the Markov chain will be unable to sample this region.
To remedy this we switch to a parameterization in terms of $\ln(\Delta
\Neff)$ and $\ln(\mster) = \ln(\Omega_{\nu, \rm sterile}h^2)+\ln(94.1)$.
This effectively scales down the proposal width of the Markov chain when
attempting to sample low $\sinsq$ and $\dm2$, giving us much greater
resolution at these small scales.
\begin{figure}[h]
\includegraphics[width=0.5\textwidth]{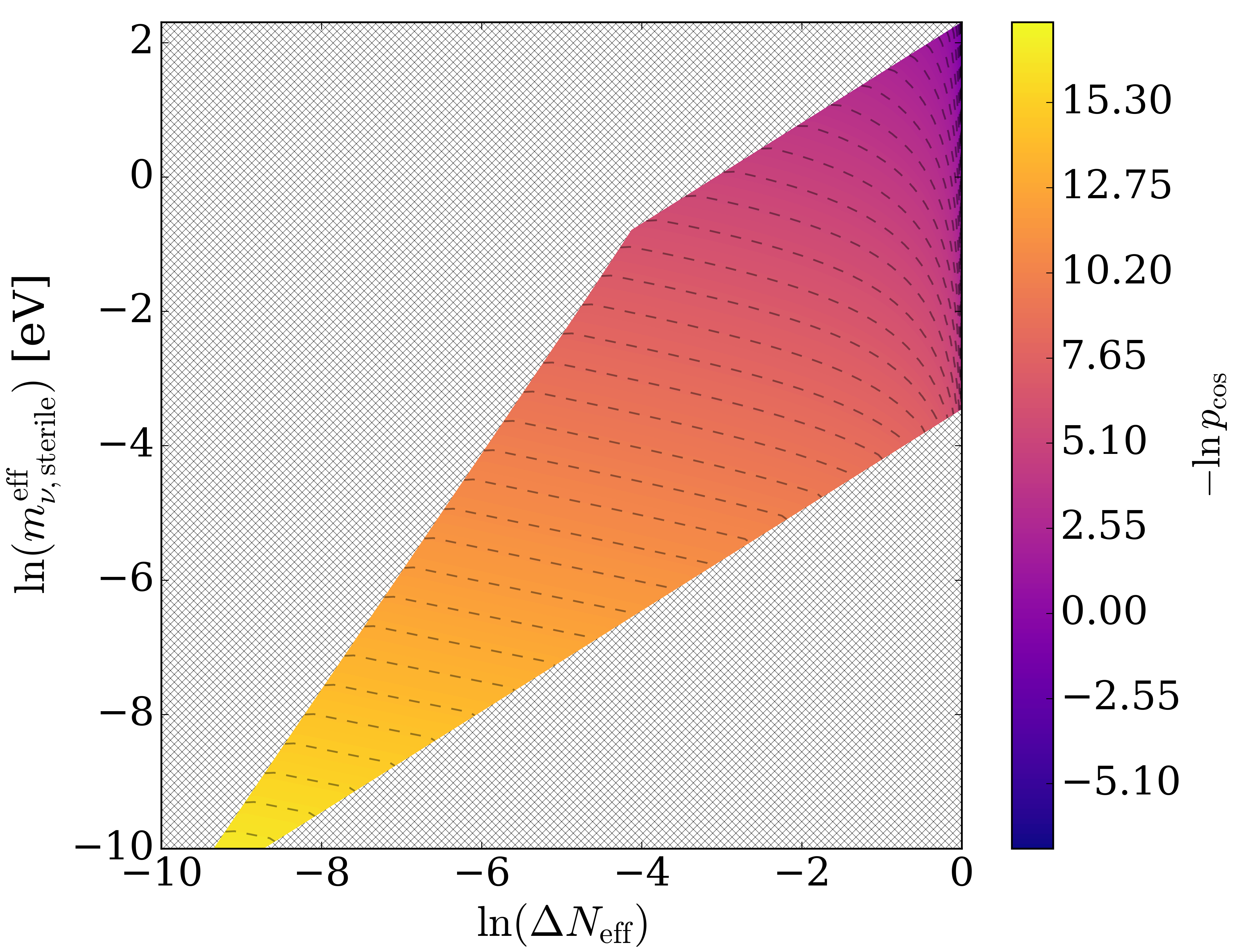}
\includegraphics[width=0.5\textwidth]{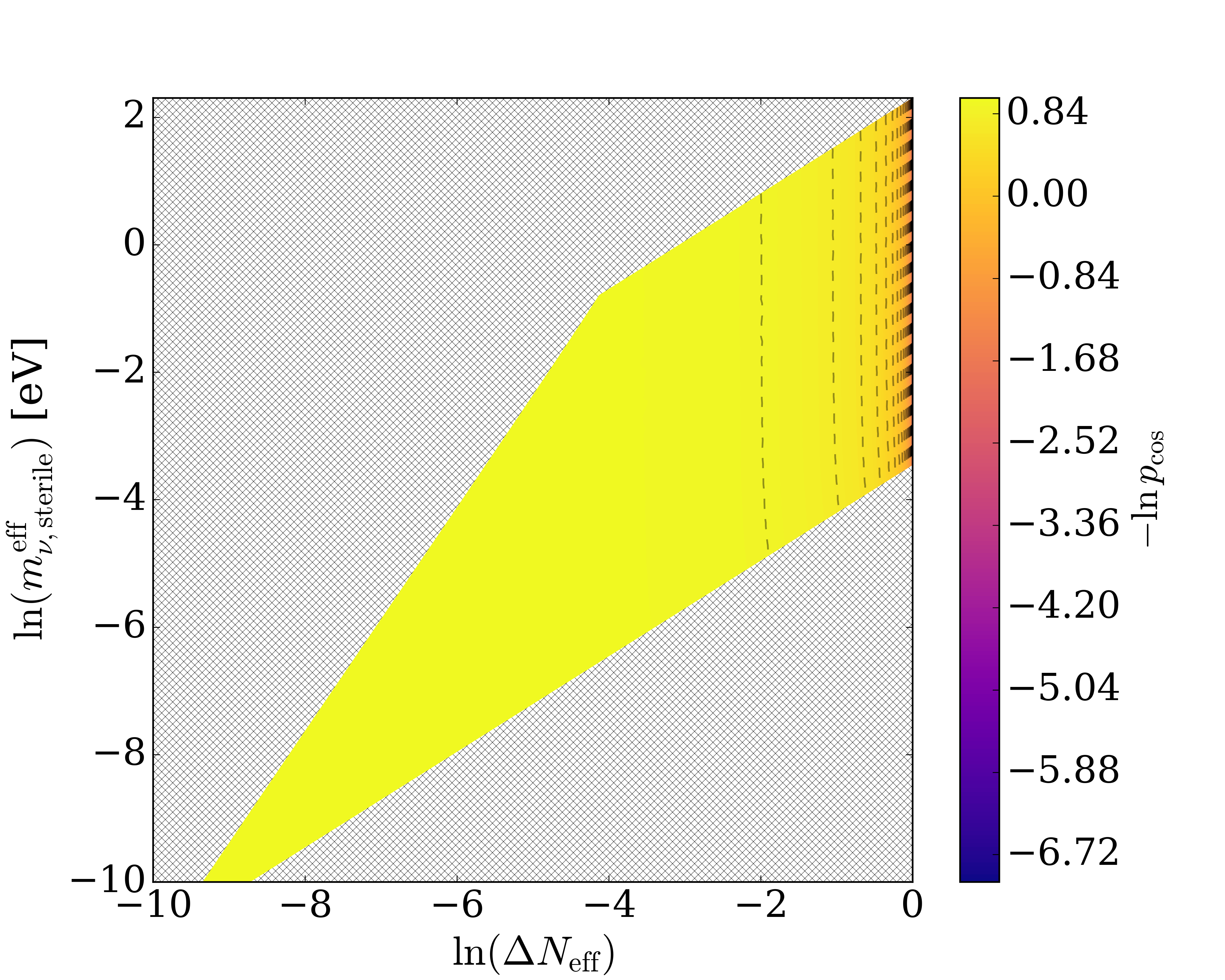}
\caption{\emph{Left panel}: Logarithmic values of the (non-normalized) Jacobian for the change of variables from the cosmology to particle parameter space, which gives flat priors in the \emph{linear} particle space, for the case where the sterile neutrino mixes with an electron neutrino. Large negative values thus correspond to larger probabilities. The probability rises sharply at large $\ln(\Delta\Neff)$ and $\ln(\mster)$, indicated by the dark blue region in the upper right corner of the plot, and so this region is favoured by the prior. The hatched regions are excluded by the chosen prior ranges for the particle parameter space discussed in section \ref{paramspaceconv}. The priors for the two different mixing cases (i.e: muon or electron) are practically identical. \emph{Right panel}: Logarithmic values of the Jacobian that gives flat priors in the \emph{log} of the particle space (i.e. log priors). This prior favours lower masses and lepton numbers more than in the left panel, since log priors weigh each order of magnitude evenly.}
\label{fig:prior}
\end{figure}
One can see in figure \ref{fig:paramspace} that the overlapping region in
the logarithmically-scaled plots is clearly visible and thus it should be
well sampled when running the MCMC. The prior in this new parameter space
that yields logarithmic priors in the particle space is
\begin{equation}
\label{eq:jac}
p_{\rm cos}\Big(\ln(\Delta\Neff), \ln(\mster)\Big) \propto \biggl|\biggl|\frac{\partial(\log(\sinsq))\,\partial(\log(\dm2))}{\partial(\ln(\Delta \Neff)\,\partial(\ln(\mster))}\biggl|\biggl|.
\end{equation}

We compute this Jacobian for each point in the cosmology parameter space
by finding the partial derivatives numerically in each direction and then
inverting the determinant of the resulting $2\times2$ matrix:
\begin{equation}
\mathbb{J}=
\begin{pmatrix}
    \dfrac{\partial(\ln(\Delta\Neff))}{\partial(\log(\sinsq))} &&& \dfrac{\partial(\ln(\Delta\Neff))}{\partial(\log(\dm2))} \\
    \dfrac{\partial(\ln(\mster))}{\partial(\log(\sinsq))} &&& \dfrac{\partial(\ln(\mster))}{\partial(\log(\dm2))} \\
\end{pmatrix}.
\end{equation}
The prior is implemented in \texttt{CosmoMC} by adding $-\ln p_{\rm cosm}$
to the negative log-likelihood function ($-\ln\mathcal{L}$) at each chain
step, thereby changing
the posteriors. We also introduce new prior ranges on the cosmology
parameters,
\begin{equation}
\label{eq:cosrange}
-10\leq\ln(\Delta\Neff)\leq0, \quad -10\leq\ln(\mster)\leq\ln{3}.
\end{equation}
Since we have switched to logarithmic space, the baseline \LCDM\ model with
$\Delta\Neff=0$ and $\mster=0$ is formally excluded; however, the
lower limit of $e^{-10}$ for both of these parameters is still very nearly standard \LCDM. Since the default MCMC sampling method in \texttt{CosmoMC} is not well
suited for dealing with priors with unusual distributions, we run
\texttt{CosmoChord} alongside \texttt{CosmoMC}; this is a nested sampling
tool designed to handle any arbitrarily complicated distribution \cite{g, h}.

The Jacobians that give uniform and log priors in the particle space are shown in figure~\ref{fig:prior}. The uniform prior on the particle parameters has the largest
probabilities at high $\Neff$. This is because, as noted in ref.~\cite{c},
large mixing angles ($\sinsq\geq0.1$) require near complete thermalization
of sterile
neutrinos, provided that there is no lepton asymmetry. A uniform prior on
the mixing angle will hence favour $\Delta\Neff \approx 1$. The sharp rise is less pronounced for log priors, owing to the fact that less weight is being placed on values of $\Delta\Neff$ that are nearly unity. 

\subsection{Choosing a prior} \label{priorchoice}

Because these priors are quite different, they will have a significant effect on the final posteriors, and so it is very important to discuss the choice of prior. Log priors exclude values of zero for a parameter, since the log of that parameter goes out to negative infinity at zero. This is in contrast to uniform priors, where zero can be included within the prior ranges. If one uses log priors, then a non-zero lower prior limit must be chosen. This can be a source of confusion, since a value of zero is a priori excluded, which might lead one to mistakenly interpret a result using that prior as a genuine detection when it is really just a function of the choice of prior. Therefore, it is important to recognize that small non-zero posteriors at the lower limits of log priors is not indicative of a detection. 

Since we want to present cosmological constraints on sterile neutrinos in a fashion that is easily comparable with constraints from particle physics experiments, we choose to implement logarithmic priors on the particle parameters in our analysis here. The mass-splitting is a scale parameter, and so the most appropriate uninformative prior in this case is one that weighs each order of magnitude uniformly, i.e. a logarithmic prior. Since $\theta$ is an angle parameter, one choice of an uninformative prior is uniform in $[0,2\pi]$. However, given that the quantity $\sinsq$ explicitly appears in the probability formula for neutrino oscillations, as shown in our earlier discussion in section \ref{sec:nucosmology}, it is also reasonable to parameterize the mixing through this parameter. Furthermore, the use of log priors will tend to avoid biasing towards a detection, since log priors place more weight on smaller scales compared to uniform priors.


\section{Results} \label{sec:results}

To summarize our approach, we perform a Markov-chain Monte Carlo analysis using \texttt{CosmoMC} to sample the posteriors for a 2-parameter
extension to the base-\LCDM\ model,
\begin{equation}
\{\Omega_{\rm b}h^2,\,\Omega_{\rm c}h^2,\,100\theta_{\rm MC},\,\tau,\,\ln(10^{10}A_{\rm s}),\,n_{\rm s},\,\ln(\Delta\Neff),\,\ln(\mster)\},
\end{equation}
and we impose a Jacobian transformation for the prior on the cosmology parameters with the prior ranges in eq.~\ref{eq:cosrange}, such that we have logarithmic priors in the particle parameter space in the ranges 
\begin{equation}
    10^{-3}\leq\dm2\leq10^{2}\,\,\rm{eV}^2, \quad 10^{-5}\leq\sinsq\leq1.
\end{equation}
The standard \LCDM\ parameters also have flat priors, as usual. The corresponding
chains for the particle parameters are calculated from the cosmology
parameters using eq.~\ref{eq:1} and by interpolating over the grid for
$\sinsq$. We also adopt the normal hierarchy, and assume that the SM neutrinos can be accurately approximated by a single massive species and two massless species, with $\Sigma m_{\nu}$ fixed at 0.06 eV. Two distinct mixing cases are considered: one where the sterile species mixes with an electron neutrino; and another where it mixes with a muon neutrino. We compare the results with MCMC runs that have flat priors in the cosmology space, as well as with the constraints from neutrino experiments. To be explicit, the data sets used in our MCMC analysis are: \\

\noindent {\bf TT+lowTEB power spectra} -- \emph{Planck} 2015 high-$\ell$
($30\leq\ell\leq2508$) CMB temperature power spectrum combined with low-$\ell$
($2\leq\ell\leq29$) temperature and LFI polarization data, which uses the \texttt{Plik} likelihood code \cite{o}; \\
\noindent {\bf Lensing} -- \emph{Planck} full-sky lensed CMB \texttt{SMICA} reconstruction \cite{p}; \\
\noindent {\bf BAO} -- baryon acoustic-oscillation data sets DR11CMASS,
DR11LOWZ, 6DF, and MGS from the SDSS-III Baryon Oscillation
Spectroscopic Survey (BOSS) \cite{l}.

\afterpage{
\begin{figure}
\centering
\includegraphics[width=1.0\textwidth]{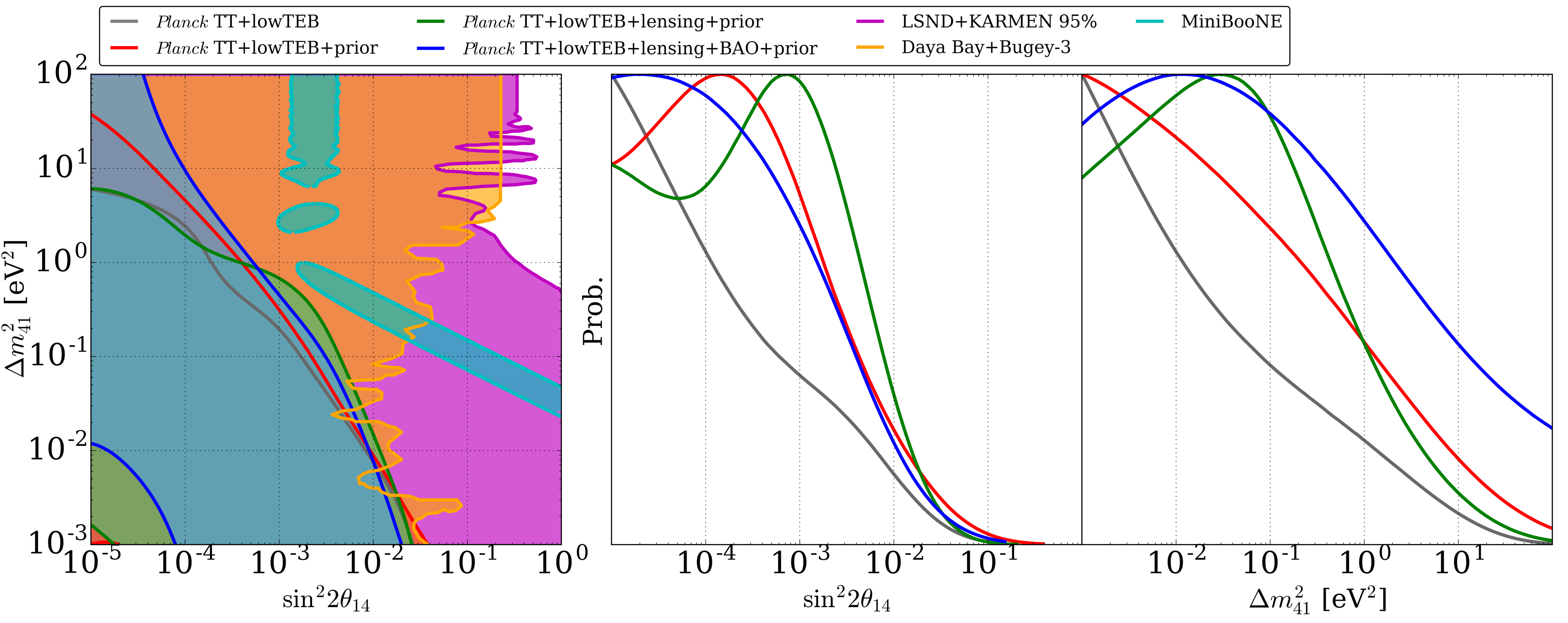}
\caption{\emph{Left panel}: Joint constraints on the particle parameters for the electron neutrino mixing case, for the two sets of priors. The shaded regions represent the 90\% confidence limits unless otherwise stated. The grey contour is the \emph{Planck} TT+lowTEB constraint with flat priors in the cosmology space, and the red contour is the \emph{Planck} TT+lowTEB constraint with log priors in the particle space. The green and blue contours are the constraints with CMB lensing and BAO likelihoods added, respectively, and log priors in the particle space. The regions not enclosed by the shaded contours are excluded. For comparison, we have included constraints from some electron neutrino and antineutrino disappearance experiments: a combined KARMEN and LSND analysis \cite{i1} (magenta); the Daya Bay/Bugey-3 sterile neutrino search \cite{h1} (orange); and the recent MiniBooNE electron neutrino and antineutrino appearance experiment \cite{b} (cyan). Note that in the legend, ``prior'' denotes logarithmic priors in the particle space. \emph{Right panels}: One-dimensional posteriors for the particle parameters from our analysis of \emph{Planck} data. The line style and colour coding follows the same scheme as in the left panel.}
\label{fig:planck_particle_14}
\vspace{0.5cm}
\includegraphics[width=1.0\textwidth]{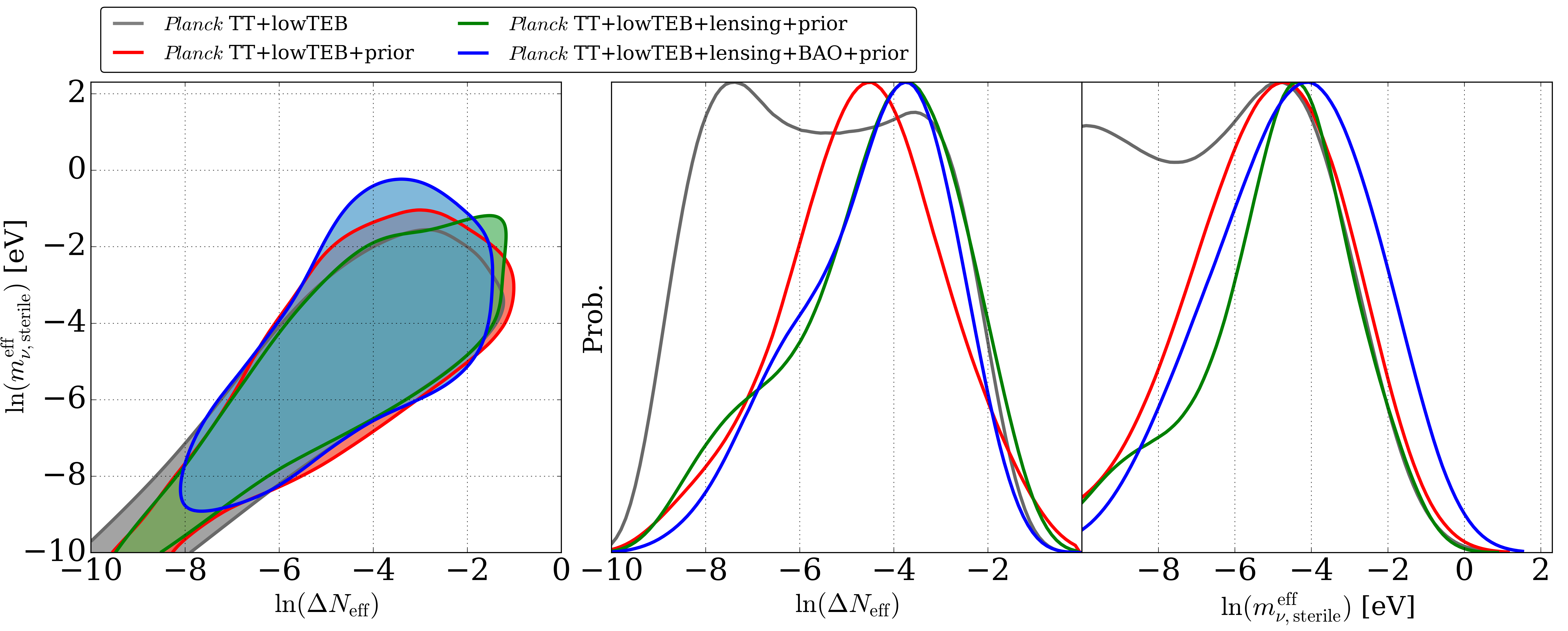}
\caption{\emph{Left panel}: Joint constraints on the cosmology parameters, following the same colour scheme as in figure \ref{fig:planck_particle_14}. All constraints are at the 90\% confidence level. \emph{Right panel}: One-dimensional posteriors for the cosmology parameters.}
\label{fig:planck_cosmo_14}
\end{figure}
\clearpage
}

\subsection{Joint constraints on $\dm2$ and $\sinsqe$} \label{sec:constraints_dm2_sinsqe}

In this section we discuss the constraints from our MCMC analysis for the
case where the sterile neutrino mixes with an electron neutrino. The top-left panel in figure \ref{fig:planck_particle_14} shows the constraints in the particle space, with and without logarithmic priors in this space. With flat priors in the cosmology space, \emph{Planck} data
rule out large mass splittings and mixing angles at over
90\% confidence. Furthermore, the posterior projections run up against the lower mass-splitting and mixing-angle limits, and are hence consistent with no sterile neutrinos. Furthermore, the region of low mixing angle and mass splitting is completely filled, demonstrating that this region of parameter space is being properly sampled by the MCMC.

Assuming flat priors in the particle
parameter space changes the constraints noticeably. The TT+lowTEB constraints on the mass gives a modest change from the case of flat priors in the cosmology space; however, there is an increased preference for larger mixing angles. When CMB lensing and BAO data added, there is a slight preference for both larger mass-splittings and larger mixing angles. In particular, when all three data sets are used, the lower left region of the particle space is not completely filled by the 90\% confidence contours. Low mixing angles are instead only favoured if the mass-splitting is above about 0.01 eV$^2$, and higher mixing angles are preferred if the mass is small. One might therefore interpret this as a possible hint of an intermediate-mass sterile neutrino with low mixing interactions with the SM neutrinos, or perhaps of a very light sterile neutrino with larger levels of mixing. The latter could be suggestive of a massless sterile neutrino, which has been considered in past cosmological studies, such as in ref.~\cite{Feng:2017nss}, in which the authors also claim to have found a hint of a massless sterile neutrino. 

The 68\% confidence limits for all relevant parameters are shown in table
\ref{tab:14results}. Below we quote the 95\% confidence limits on the particle parameters:
\begin{align}
\begin{rcases}
\dm2 < 5.9\,\,{\rm eV}^2, \\
\sinsqe < 0.011, 
\end{rcases}
&{\rm 95\%\,\,TT\hspace{-0.1cm}+\hspace{-0.1cm}lowTEB\hspace{-0.1cm}+\hspace{-0.1cm}prior}, \\
\begin{rcases}
\dm2 < 2.6\,\,{\rm eV}^2, \\
\sinsqe < 0.0089, 
\end{rcases}
&{\rm 95\%\,\,TT\hspace{-0.1cm}+\hspace{-0.1cm}lowTEB\hspace{-0.1cm}+\hspace{-0.1cm}lensing\hspace{-0.1cm}+\hspace{-0.1cm}prior}, \\
\begin{rcases}
\dm2\,\,{\rm unconstrained}, \\
\sinsqe < 0.0068,
\end{rcases}
&{\rm 95\%\,\,TT\hspace{-0.1cm}+\hspace{-0.1cm}lowTEB\hspace{-0.1cm}+\hspace{-0.1cm}lensing\hspace{-0.1cm}+\hspace{-0.1cm}BAO\hspace{-0.1cm}+\hspace{-0.1cm}prior.}
\end{align}
With logarithmic priors in the particle space, \emph{Planck} TT+lowTEB power spectrum data yield upper bounds of $\dm2<5.9$ eV$^2$ and $\sinsqe < 0.011$ at the 95\% confidence level. CMB lensing brings the mixing angle constraint down to $\sinsqe < 0.0089$. The addition of CMB lensing and BAO further tightens the mixing angle constraint, but BAO weakens the constraint on the mass-splitting. In fact, BAO increases the likelihood at larger mass-splittings, compared to when BAO are not included, as seen by the posteriors in the right panel of figure \ref{fig:planck_particle_14}. However, we emphasize that the 1D posteriors presented in our analysis are still highly non-zero at the lower prior limits. As noted in our discussion of priors in section \ref{priorchoice}, the use of log priors technically means that zero mass-splitting and zero mixing are excluded by the prior ranges. In this case, non-zero posteriors at the lower limit of log priors are typically interpreted as plateaus that extend towards negative infinity. Our results here are therefore not indicative of any detection, and we conclude that the \emph{Planck} data remain consistent with the scenario of no sterile neutrinos.

In figure \ref{fig:planck_cosmo_14} we have shown the constraints on the cosmology space. With flat priors in the
log-cosmology space, the posteriors run up against the lower prior edges,
consistent with the grey contours in figure \ref{fig:planck_cosmo_14}. Switching to flat priors in the particle parameter space leads to slightly higher likelihoods for larger effective neutrino number and effective masses, and this is reflected in the particle constraints as increased likelihoods for higher mass-splittings. For TT+lowTEB+lensing+BAO, the 90\% contours no longer run up against the lower prior limit, and so the apparent preference for higher mass-splittings is also visible in the cosmology parameters. This also provides a clearer picture of why BAO data weaken the constraint on the mass splitting. BAO lowers the likelihoods at low $\Delta\Neff$ and $\mster$, while increasing the likelihood at high $\mster$, but leaving the mixing-angle posterior roughly the same as TT+lowTEB+lensing near the peak. This results in a better mixing-angle constraint, but a weaker mass-splitting constraint. The addition of lensing and BAO data to the \emph{Planck} TT+lowTEB likelihood gives the highest value on the effective mass, and this is also where we see the highest upper bound on the mass splitting. These
constraints nonetheless fall well within the $(\Neff, \mster)$ confidence
limits of the Planck Collaboration analysis \cite{c}. Fully thermalized
sterile neutrinos with $\Delta\Neff=1$ remain excluded at 90\% confidence for all likelihood combinations. This suggests
that any sterile neutrinos would have to be incompletely thermalized in the
early Universe through some sort of cosmological suppression mechanism. 
\begin{table}
\centering
\fontsize{10}{10}\selectfont
\begin{tabular}{llll}
\hline
\hline
\multicolumn{1}{C{1cm}}{Parameter} &
\multicolumn{1}{C{3cm}}{TT+lowTEB+prior 68\% limits} &
\multicolumn{1}{C{3cm}}{TT+lowTEB +lensing+prior\hspace{2cm}68\% limits} &
\multicolumn{1}{C{3cm}}{TT+lowTEB +lensing +BAO+prior\hspace{2cm}68\% limits} \Tstrut \Bstrut \\
\hline
$\Omega_{\rm b}h^2$ & $0.02221\pm 0.00024$ & $0.02230\pm 0.00022$ & $0.02229\pm 0.00020$ \Tstrut \\
$\Omega_{\rm c}h^2$ & $0.1203\pm 0.0023$ & $0.1188\pm 0.0023$ & $0.1183\pm 0.0017$ \\
$100\theta_{\rm MC}$ & $1.04081\pm 0.00047$ & $1.04099\pm 0.00047$ & $1.04097\pm 0.00041$ \\
$\tau$ & $0.084\pm 0.021$ & $0.069\pm 0.016$ & $0.068\pm 0.014$ \\
$\ln(10^{10}A_{\rm s})$ & $3.095\pm 0.035$ & $3.069\pm 0.029$ & $3.068\pm 0.026$ \\
$n_{\rm s}$ & $0.9654\pm 0.0068$ & $0.9694\pm 0.0060$ & $0.9687\pm 0.0051$ \Bstrut \\
\hline
$\Delta\Neff$ & $0.010_{-0.0076}^{+0.069}$ & $0.01_{-0.0091}^{+0.14}$ & $0.01_{-0.0088}^{+0.10}$  \Tstrut \\
$\mster$ [eV] & $0.007_{-0.0057}^{+0.067}$ & $0.008_{-0.0061}^{+0.084}$ & $0.01_{-0.010}^{+0.13}$ \\
$\dm2$ [eV$^2$] & $< 0.14$ & $< 0.26$ & $< 0.41$  \\
$\sinsqe$ & $0.00026_{-0.00025}^{+0.00098}$ & $< 0.0011$ & $< 0.00056$ \\
$H_0$ [km s$^{-1}$ Mpc$^{-1}$] & $67.3\pm 1.1$ & $68.00\pm 0.95$ & $67.94\pm 0.61$\\
$\sigma_8$ & $0.828\pm 0.016$ & $0.812\pm 0.012$ & $0.810\pm 0.015$ \Bstrut \\
\hline
\end{tabular}
\caption{68\% CL parameter constraints on the 2-parameter extension to the
base-\LCDM\ model for the electron-neutrino-mixing case, along with relevant derived parameters. The cosmology parameter constraints have been converted to linear values for ease of interpretation. Parameters with nearly Gaussian distributions are written with their mean and standard deviation, whereas non-Gaussian parameters have their mean and 68\% limits quoted. For unconstrained parameters, only the 68\% bounds are shown.}

\label{tab:14results}
\end{table}
\begin{figure}
\centering
\includegraphics[width=0.85\textwidth]{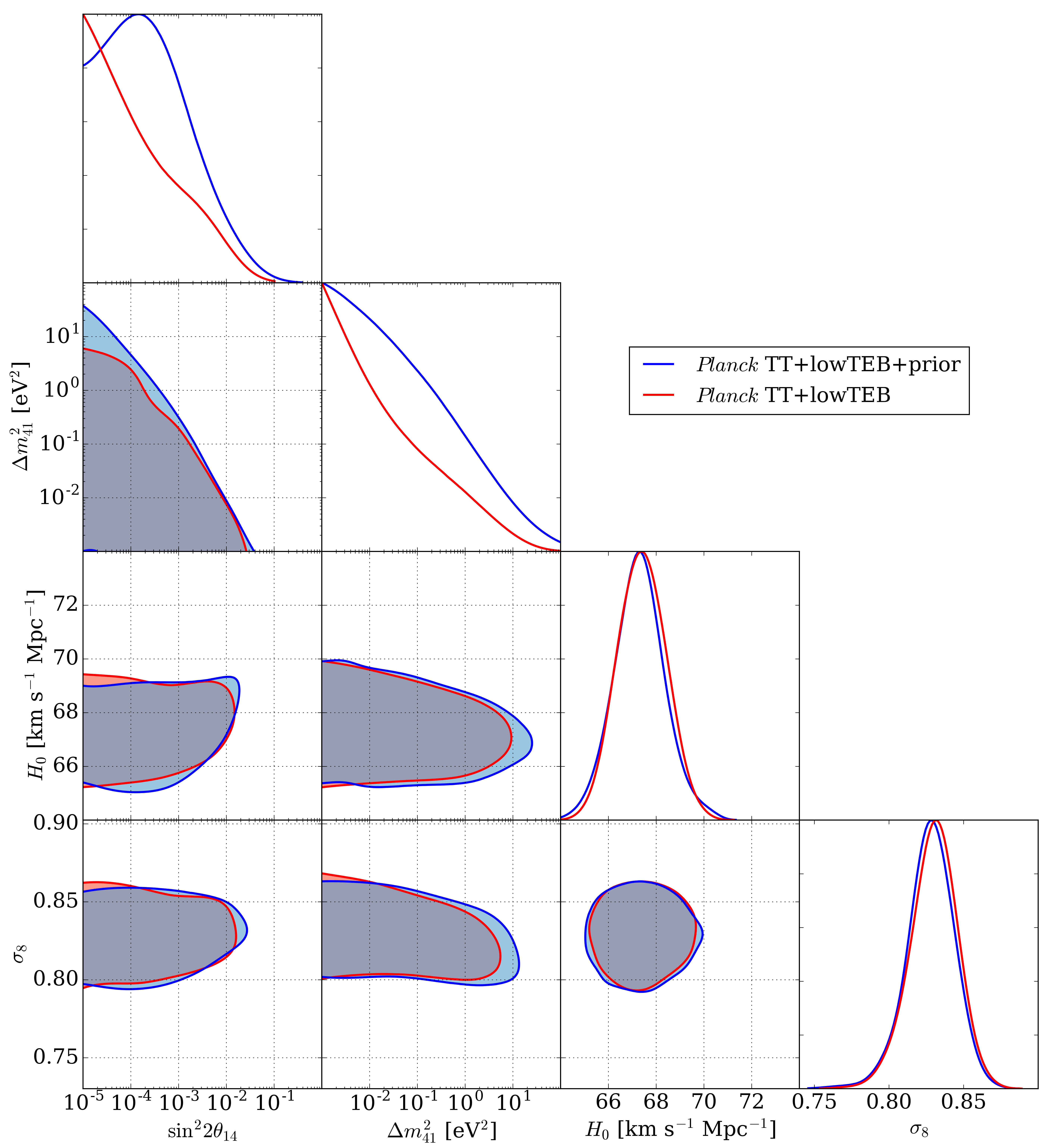}
\caption{Covariances between the particle parameters, $H_0$, and $\sigma_8$ for
the electron-neutrino mixing case. The solid blue lines are the results with
flat priors in the particle space, and the dashed red lines are for flat priors
in the log-cosmology space. The \emph{Planck} TT+lowTEB data set is used here. Limits are at 90\% confidence.}
\label{fig:planck_tri14}
\end{figure}

Because we have to explore the region of parameter space corresponding to very small $\Delta\Neff$ and $\mster$ down to the order of $10^{-5}$ to properly sample the particle parameters, \texttt{LASAGNA} enters scales at which it begins to encounter issues with accuracy. In the low mass-splitting region near $\dm2 \sim 10^{-3}\,{\rm eV}^2$, the error in $\Neff$ is as large as 0.01 (see ref.~\cite{Hannestad:2015tea} for a discussion of the accuracy of \texttt{LASAGNA}). Hence, the reader should note that the output of \texttt{LASAGNA} at the lower corner of the parameter space in figure \ref{fig:planck_cosmo_14} is somewhat unreliable.

The \emph{Planck} constraints are largely consistent with the neutrino disappearance searches shown in figure \ref{fig:planck_particle_14}. There is mild tension with Daya Bay+Bugey-3 at low mass-splitting, but given that these constraints are derived from completely different sets of data, some tension is to be expected. The cosmological constraints are much stronger than Daya Bay+Bugey-3 and LSND+KARMEN, excluding a much larger region of the parameter space. The recent detection of an excess of electron neutrino events by the MiniBooNE experiment has received a great deal of attention, and so we have also included these constraints in our plots. The MiniBooNE limits are in very strong tension with cosmology, as well as with various other particle physics experiments. The constraints from cosmology in particular completely rule out the regions allowed by MiniBooNE and LSND. However, we reiterate what we stated in section~\ref{sec:nucosmology}, that $\nu_\mu \rightarrow \nu_e$ appearance channels require a two-flavour oscillation model, and we have only constrained single flavour oscillation models, so these sets of constraints are not necessarily directly comparable. 

Since a great deal of attention has been give to possible tensions between CMB
measurements and several low-redshift parameter constraints \cite{j1}, it is worth
examining how sterile neutrinos might affect these tensions. In figure
\ref{fig:planck_tri14} we have plotted the covariances between the particle parameters and the linear power spectrum amplitude $\sigma_8$, and the Hubble parameter $H_0$. Increasing the mixing angle or the mass-splitting has no significant affect on the value of $\sigma_8$. Larger mixing angles lead to marginal increases in the value of $H_0$, and so high sterile-neutrino mixing may provide a path towards resolving the $H_0$ tensions. There is little correlation between the mass-splitting and $H_0$, but increasing the mass does appear to lower $\sigma_8$ slightly. The switch to log priors does not affect the distributions of $\sigma_8$ and $H_0$ significantly either.

In our earlier discussion of priors, we noted that a log prior must have a non-zero lower limit. It therefore seems plausible that constraints derived with such priors may be sensitive to the choice of this lower limit, and in the case of sterile neutrinos, choosing too high of a lower limit could mask potential signatures of a detection. We ran additional MCMC chains using a lower limit on the mass splitting of $10^{-5}$ eV$^2$ to check the robustness of our constraints, finding no significant differences that would affect our overall conclusions.

\subsection{Joint constraints on $\dm2$ and $\sinsqmt$} \label{sec:constraints_dm2_sinsqmt}

Having the sterile neutrino mix with a muon neutrino,
as opposed to an electron neutrino, makes very little difference to the
calculation of $\Delta\Neff$ and $\mster$. The constraint on $\sinsqmt$ will be
very similar to the constraint on $\sinsqe$, and thus most of what has been said
about the latter will be applicable to the former. Since the Jacobians for the
two cases are quite similar, the ratio of the Jacobians is approximately
constant, and so we can obtain the corresponding posteriors for the muon
neutrino mixing case by multiplying the likelihoods from the previous set of chains by this ratio.
\begin{figure}
\centering
\includegraphics[width=1.0\textwidth]{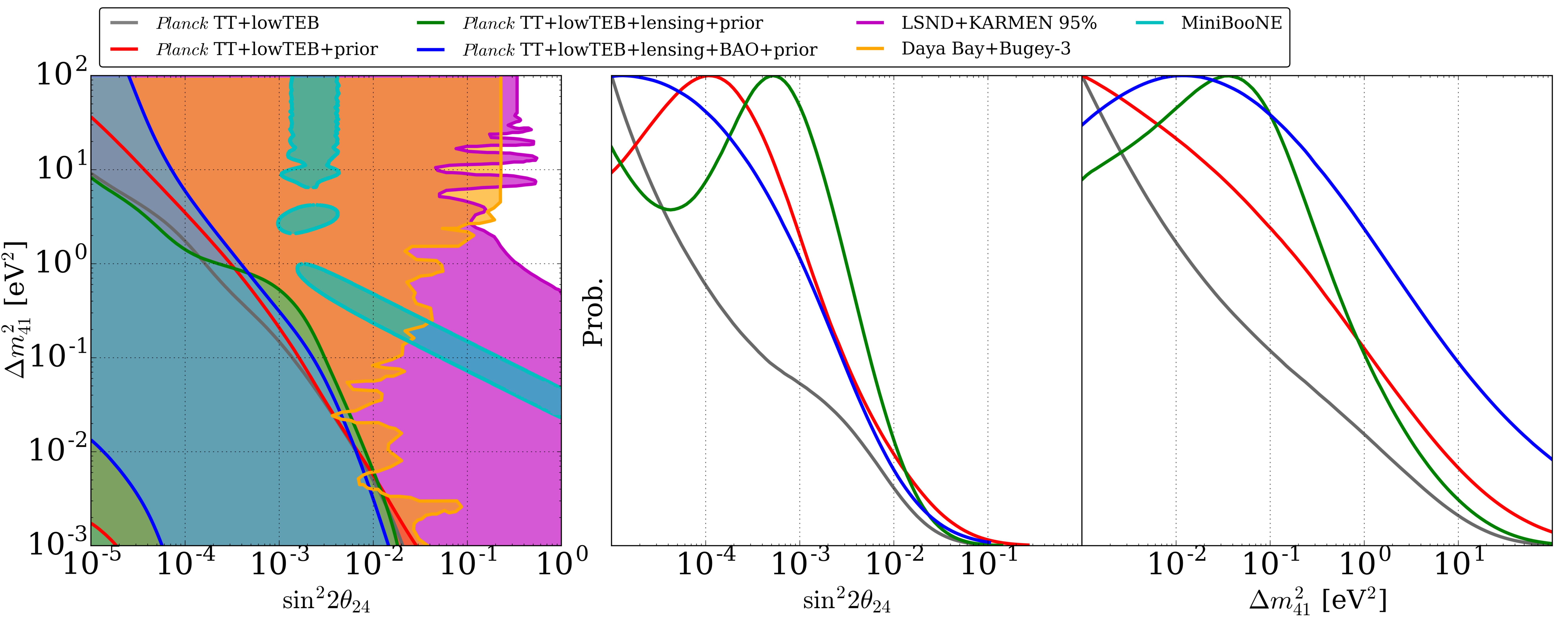}
\caption{\emph{Left panel}: Joint constraints on the particle parameters for the muon-neutrino mixing case, for the two sets of priors. The shaded regions represent the 90\% confidence limits unless otherwise stated. The colour scheme is the same as in figure \ref{fig:planck_particle_14}. Note that in the legend, ``prior'' denotes logarithmic priors in the particle space. \emph{Right panels}: One-dimensional posteriors for the particle parameters from our analysis of \emph{Planck} data. The line style and colour coding follows the same scheme as in the left panel.}
\label{fig:planck_particle_24}
\end{figure}
The 95\% limits for the particle parameters for the muon neutrino mixing case are:
\begin{align}
\begin{rcases}
\dm2 < 5.1\,\,{\rm eV}^2, \\
\sinsqmt < 0.0088, 
\end{rcases}
&{\rm 95\%\,\,TT\hspace{-0.1cm}+\hspace{-0.1cm}lowTEB\hspace{-0.1cm}+\hspace{-0.1cm}prior};
\end{align}
\begin{align}
\begin{rcases}
\dm2 < 2.4\,\,{\rm eV}^2, \\
\sinsqmt < 0.0067, 
\end{rcases}
&{\rm 95\%\,\,TT\hspace{-0.1cm}+\hspace{-0.1cm}lowTEB\hspace{-0.1cm}+\hspace{-0.1cm}lensing\hspace{-0.1cm}+\hspace{-0.1cm}prior}; \\
\begin{rcases}
\dm2\,\,{\rm unconstrained}, \\
\sinsqmt < 0.0052,
\end{rcases}
&{\rm 95\%\,\,TT\hspace{-0.1cm}+\hspace{-0.1cm}lowTEB\hspace{-0.1cm}+\hspace{-0.1cm}lensing\hspace{-0.1cm}+\hspace{-0.1cm}BAO\hspace{-0.1cm}+\hspace{-0.1cm}prior};
\end{align}
and the contours are plotted in figure~\ref{fig:planck_particle_24}.

\section{Forecasting}
\label{sec:forecast}

Several future CMB experiments, like CMB-S4 \cite{a1, b1}, aim to improve the
sensitivity to cosmological neutrinos, and therefore find more stringent constraints on sterile neutrinos than \emph{Planck}. Fisher-matrix methods are commonly used
by cosmologists to predict the precision with which these future experiments
could measure cosmological parameters, and in this section we proceed to apply
these techniques to forecast sterile neutrino constraints in the particle
parameter space.

\subsection{The CMB Fisher matrix}

To forecast constraints from the CMB, one must first calculate the Fisher
matrix for a given fiducial (baseline) model. For an $n$-parameter model, the
Fisher matrix is a square $n\times n$ matrix that fully encodes all the
information from the CMB. Calculating Fisher matrices normally involves finding
the first moment of the Hessian of the log-likelihood function for the fiducial
model. However, by assuming Gaussian perturbations in the CMB temperature and
polarization anisotropies, the Fisher matrix for the CMB can be rewritten as
\cite{z}
\begin{equation} \label{eq:fisher}
F_{ij} = \sum_\ell \sum_{XY} \dfrac{\partial C_\ell^X}{\partial p_i}(\mathbb{C})_{XY}^{-1} \dfrac{\partial C_\ell^Y}{\partial p_j}
\end{equation}
where $X=T,E,C$, respectively, denote the temperature, $E$-mode
polarization, and temperature-polarization cross-correlation terms, $p_i$ are
the model parameters, and $\mathbb{C}$ is the CMB covariance matrix
\begin{equation}
\mathbb{C}=
\begin{pmatrix}
    (\mathbb{C})_{TT} && (\mathbb{C})_{TE} && (\mathbb{C})_{TC} \\
    (\mathbb{C})_{TE} && (\mathbb{C})_{EE} && (\mathbb{C})_{EC} \\
    (\mathbb{C})_{TC} && (\mathbb{C})_{EC} && (\mathbb{C})_{CC}
\end{pmatrix}.
\end{equation}
Here we have ignored the primordial $B$-modes, since they are yet to be well
measured and are likely far too weak for our model to be sensitive to them
\cite{c1}. The Cramer-Rao inequality then places upper bounds on the errors for
the model parameters, $\sigma_i\leq\sqrt{F_{ii}^{-1}}$. The inverse of the
Fisher matrix is used, since all the model parameters are allowed to float. If
one does not use the inverse here, then the errors correspond to the case where
all the other parameters are fixed. The derivatives of the $C_\ell$s in
eq.~\ref{eq:fisher} are taken about the chosen fiducial model. 

Accounting for
instrumental noise, the elements of the covariance matrix are \cite{z}
\begin{align}
  (\mathbb{C})_{TT} &= \frac{2}{(2\ell+1)f_{\rm sky}}(C_\ell^{TT}+w_T^{-1}B_\ell^{-2})^2, \\
  (\mathbb{C})_{TE} &= \frac{2}{(2\ell+1)f_{\rm sky}}(C_\ell^{TE})^2, \\
  (\mathbb{C})_{TC} &= \frac{2}{(2\ell+1)f_{\rm sky}}C_\ell^{TE}(C_\ell^{TT}+w_T^{-1}B_\ell^{-2}), \\
  (\mathbb{C})_{EE} &= \frac{2}{(2\ell+1)f_{\rm sky}}(C_\ell^{EE}+w_P^{-1}B_\ell^{-2})^2, \\
  (\mathbb{C})_{EC} &= \frac{2}{(2\ell+1)f_{\rm sky}}C_\ell^{TE}(C_\ell^{EE}+w_P^{-1}B_\ell^{-2}), \\
  (\mathbb{C})_{CC} &= \frac{1}{(2\ell+1)f_{\rm sky}}[(C_\ell^{TE})^2+(C_\ell^{TT}+w_T^{-1}B_\ell^{-2})(C_\ell^{EE}+w_P^{-1}B_\ell^{-2})],
\end{align}
The constants $w_T$ and $w_P$ are the inverse squares of the detector noise
level over a steradian patch of the sky for $T$ and $E$, $f_{\rm sky}$ is the
fraction of the sky being observed, and
\begin{equation}
  B_\ell^2 = \exp\bigg(\frac{-\ell(\ell+1)\theta_{\rm beam}^2}{8\ln2}\bigg)
\end{equation}
is the beam window function, where $\theta_{\rm beam}$ is the full-width,
half-maximum beam angle \cite{z}. The noise terms vanish in the limit of zero
detector noise, $w_T, w_P \rightarrow \infty$, yielding the covariance matrix
for the noise-free case. The elements of the Fisher matrix encode the confidence ellipses for every pair of parameters, marginalized over all other parameters. Given two parameters $x$ and $y$, with $1\sigma$ Fisher uncertainties $\sigma_x$ and $\sigma_y$, and with correlation cross-terms $\sigma_{xy}=\sigma_{yx}$, the ellipse axis parameters are calculated as \cite{Coe09}
\begin{align}
&a^2 = \frac{\sigma_x^2+\sigma_y^2}{2}+\sqrt{\frac{(\sigma_x^2-\sigma_y^2)^2}{4}+\sigma_{xy}^2}, \\
&b^2 = \frac{\sigma_x^2+\sigma_y^2}{2}-\sqrt{\frac{(\sigma_x^2-\sigma_y^2)^2}{4}+\sigma_{xy}^2},
\end{align}
where $a$ corresponds to the semimajor axis, and $b$ to the semiminor axis. The angle that controls the (counterclockwise) tilt of the ellipse is given by \cite{Coe09}
\begin{equation}
\tan 2\theta = \frac{2\sigma_{xy}}{\sigma_x^2-\sigma_y^2}.
\end{equation}

\begin{figure}
\centering
\includegraphics[width=0.8\textwidth]{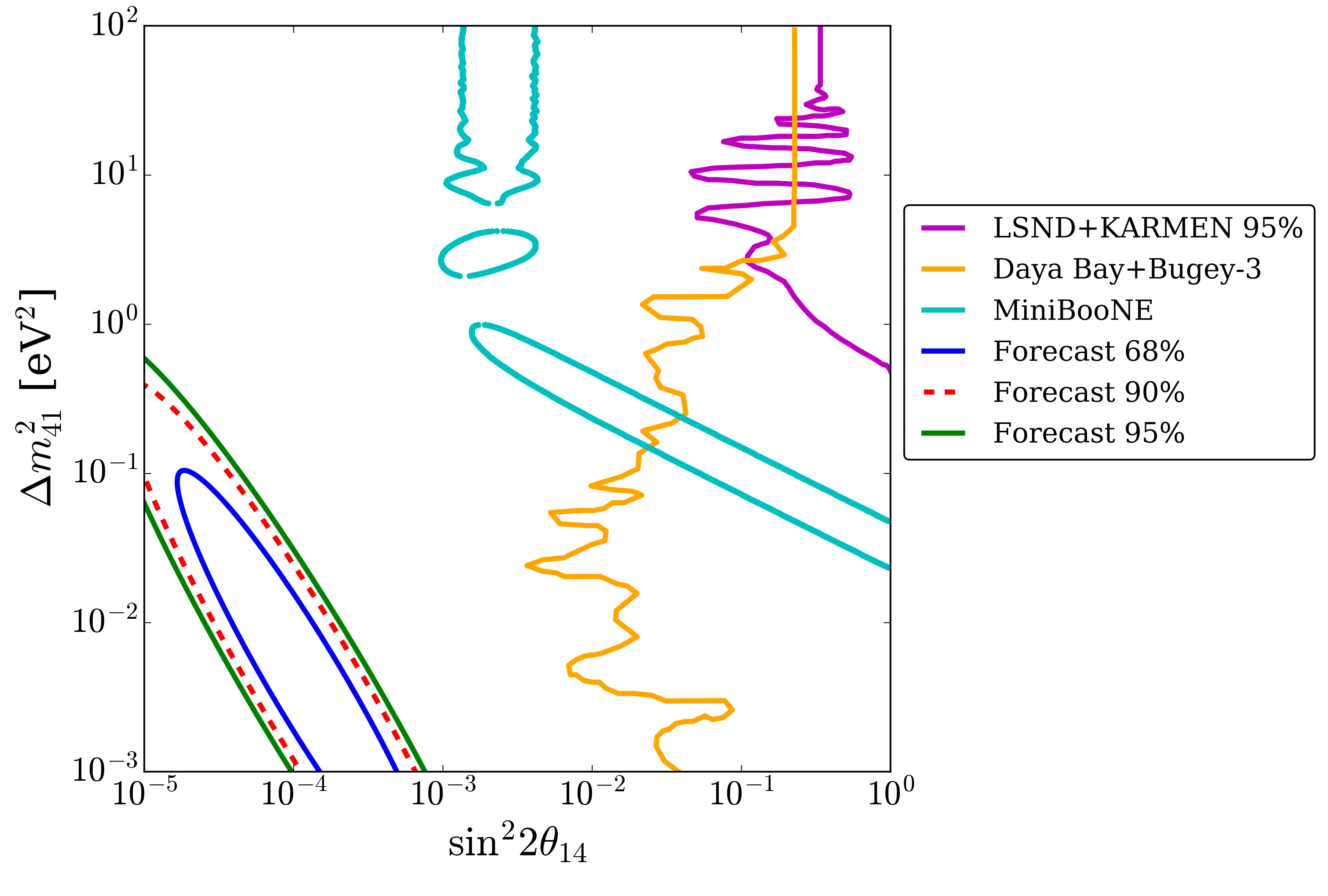}
\caption{Fisher confidence ellipses in the particle space for an experiment resembling CMB-S4, with $f_{\rm sky}=0.4$, a 1-arcmin
beam, and noise levels in $T$ and $E$ of $1\,\mu$K-arcmin and
$\sqrt{2}\,\mu$K-arcmin, respectively \cite{a1}. The covariance matrix for the full $TT+TE+EE$ spectrum is used. The fiducial model is the best-fit TT+lowTEB point, $(\dm2, \sinsqe)=(0.0038\,\,{\rm eV}^2, 0.00013)$. The contours are drawn at 68\%, 90\%, and 95\% confidence. For comparison, we have included the LSND+KARMEN 95\%, Daya Bay+Bugey-3 90\%, and MiniBooNE 90\% constraints previously shown in figures \ref{fig:planck_particle_14} and \ref{fig:planck_particle_24}. Note that the regions to the right of the orange and magenta curves are excluded.}
\label{fig:fisher}
\end{figure}

\subsection{Particle parameter forecasts}

We use \texttt{CAMB} \cite{y} to compute the power spectra in order to find the
derivatives of the $C_\ell$s with respect to each parameter for the 2-parameter extension to the \LCDM\
presented in section~\ref{sec:results}. The derivatives with respect to the log of the
particle parameters are computed from the derivatives with respect to
$\ln(\Delta\Neff)$ and $\ln(\mster)$ and then applying the chain rule. We then
calculate the Fisher matrix for an experiment with noise levels and beam
settings similar to those of CMB-S4, using the full $TT+TE+EE$ covariance matrix, which we then invert to find the uncertainties. The fiducial model is the best-fit (maximum likelihood) point from the TT+lowTEB MCMC run, $(\dm2, \sinsqe)=(0.0038\,\,{\rm eV}^2, 0.00013)$. The confidence ellipses are plotted in figure \ref{fig:fisher}. The $1\sigma$ errors on the particle parameters, in log space, are
\begin{equation}
\sigma(\log(\dm2))=0.95\,\,{\rm eV}^2, \quad \sigma(\log(\sinsqe))=0.58,
\end{equation}
and represent close to the best possible measurement on these parameters obtainable using the CMB. Figure \ref{fig:fisher} shows that this best possible CMB constraint is significantly better than what current \emph{Planck} data can do, and remains competitive with particle physics experiments. Therefore, future CMB experiments will still be of great importance with regards to further constraining sterile neutrinos, and will potentially show whether any of the hints of sterile neutrinos encountered thus far turn out to be signs of new physics or not.

\section{Conclusions}
\label{sec:conclusions}

We have used \emph{Planck} data to obtain cosmological
constraints on the sterile-neutrino oscillation parameters: the squared mass
splitting $\dm2$; and the mixing angles $\sinsqe$ and $\sinsqmt$, corresponding
to two different models where a single sterile neutrino species mixes with an
electron neutrino or a muon neutrino. The posteriors were inferred using an MCMC
analysis of \emph{Planck} CMB power spectra, lensing, and BAO data, where our
model consists of the six \LCDM\ parameters plus $\ln(\Delta\Neff)$ and
$\ln(\mster)$, which were used to vary $\dm2$ and
$\sinsq$ by proxy. We compared results that have flat
priors in the cosmology space with results that had logarithmic priors in the
particle parameter space, which was accomplished by imposing the Jacobian of
the change of variables transformation between the two parameter spaces as the
prior on the cosmology space. The \emph{Planck} data show slightly increased
preference for non-zero mass-splitting when priors are flat in the particle
space in the range $10^{-3}\leq \dm2 \leq10^2$ eV$^2$,
$10^{-5}\leq\sinsq\leq1$. For the model where the sterile neutrino mixes with an electron neutrino, we find constraints of $\dm2 < 0.41$ eV$^2$ and $\sinsqe < 0.00056$
at 68\% confidence for the TT+lowTEB+lensing+BAO likelihoods. At 95\% confidence, the mixing angle constraint increases to $\sinsqe < 0.0068$, and the probabilities for higher values of $\dm2$ also increase, but the data are unable to constrain the mass splitting at this confidence level. For the second case where the sterile neutrino mixes with a muon neutrino, we find that $\dm2$ is again unconstrained and $\sinsqmt<0.0052$ at 95\% confidence, using all three data sets. In summary, we conclude that the \emph{Planck} data indicate no evidence for sterile neutrinos. We have also compared these results with sterile neutrino constraints from particle physics experiments, and have shown that cosmology strongly rules out the region of particle space allowed by LSND and MiniBooNE. However, our analysis has assumed that sterile neutrinos mix with a single neutrino flavour, and so the assumptions made here are not exactly the same as those used by LSND and MiniBooNE.

The search for sterile neutrinos is certainly not over. The anomalies observed in LSND and MiniBooNE remain unexplained, and future particle physics experiments will no doubt attempt to hone in on the culprit, be it sterile neutrinos or something else. Future cosmology experiments also promise to find more stringent constraints on cosmological neutrinos, and will likely elaborate more on the status of sterile neutrinos within cosmology. 

\acknowledgments
This work was supported by the Natural Sciences and Engineering Research
Council of Canada (\url{www.nserc-crsng.gc.ca}). Computing
resources were provided by WestGrid (\url{www.westgrid.ca}) and Compute Canada/Calcul Canada (\url{www.computecanada.ca}). Our main results were obtained using the codes \texttt{LASAGNA}, \texttt{CAMB}, and \texttt{CosmoMC}. We also thank Julien Lesgourgues for helpful discussions on neutrino physics.

\bibliographystyle{JHEP}
\bibliography{biblio}


\end{document}